# On-demand reversible switching of the emission mode of individual semiconductor quantum emitters using plasmonic metasurfaces

Adam Olejniczak,[1] Zuzanna Lawera,[1] Mario Zapata-Herrera,[1] Andrey Chuvilin,[2,3] Pavel Samokhvalov,[4,5] Igor Nabiev,[4,5,6] Marek Grzelczak,[1,7] Yury Rakovich,[1,2,7,8,*], Victor Krivenkov[1,8,**]

[1]Centro de Física de Materiales (MPC, CSIC-UPV/EHU), Donostia - San Sebastián, 20018, Spain

[2] Ikerbasque, Basque Foundation for Science, Bilbao, 48013, Spain

[3]CIC NanoGUNE Consolider, Tolosa Hiribidea 76, Donostia - San Sebastián 20018, Spain

[4]Life Improvement by Future Technologies (LIFT) Center, Skolkovo, 143025 Moscow, Russia

[5]National Research Nuclear University MEPhI (Moscow Engineering Physics Institute), 115409 Moscow, Russia

[6]Laboratoire de Recherche en Nanosciences, LRN-EA4682, Université de Reims Champagne-Ardenne, 51100 Reims, France

[7]Donostia International Physics Center (DIPC), Donostia - San Sebastián, 20018, Spain

[8]Polymers and Materials: Physics, Chemistry and Technology, Chemistry Faculty, University of the Basque Country (UPV/EHU), Donostia - San Sebastián, 20018, Spain

* yury.rakovich@ehu.eus (YR) and **victor.krivenkov@ehu.eus (VK).

**Abstract**

The field of quantum technology has been rapidly expanding in the past decades, yielding numerous applications as quantum information, quantum communication and quantum cybersecurity. The central building block for these applications is a quantum emitter (QE), a controllable source of single photons or photon pairs. Semiconductor QEs such as perovskite nanocrystals (PNCs) and semiconductor quantum dots (QDs) have been demonstrated to be a promising material for pure single-photon emission, and their hybrids with plasmonic nanocavities may serve as sources of photon pairs. Here we have designed a system in which individual quantum emitters and their ensembles can be traced before, during, and after the interaction with the external plasmonic metasurface in controllable way. Upon coupling the external plasmonic metasurface to the array of QEs, the individual QEs switch from single-photon to photon-pair emission mode. Remarkably, this method does not affect the chemical structure and composition of the QEs, allowing them to return to their initial state after decoupling from the plasmonic metasurface. By employing this approach, we have successfully demonstrated the reversible switching of the ensemble of individual semiconductor QEs between single-photon and photon pair emission modes. This significantly broadens the potential applications of semiconductor QEs in quantum technologies.

**Introduction**

The second quantum revolution will affect many sectors of society, including healthcare, finance, defense, weather modeling, and cybersecurity.[1] It has given rise to the advanced areas of quantum communication and computing, collectively termed quantum information, i.e., the transmission, storage, and processing of information using quantum systems. The basic unit of quantum information is a quantum bit (qubit), which is fundamentally different from the classical bit, and most of current quantum information protocols use photon-based qubits. Sources of light used in quantum information, called quantum emitters (QEs), must have specific photon statistics. The following characteristics are the most important for an ideal QE: a high single-photon purity, indistinguishability, a high operation rate and a high brightness.[2] Some of the advanced quantum information protocols include quantum entangled photon pair.[3,4] On-demand sources of photon pairs are also referred to as QEs.[5,6] Indistinguishable single photons and entangled photon pairs are routinely generated by parametric down-conversion.[7]

However, this nonlinear process gives a probabilistic amount of photons, which also implies the generation of zero and multiple photons, and cannot produce single photon or photon pair on demand. Thus, it cannot be used as QEs, and recent efforts in designing QEs are focused on cold atoms and ions, superconducting circuits, and solid-state sources, such as emitting vacancy centers in diamond or fluorescent nanocrystals.[8,9] Along with others, solid-state QEs are more flexible in terms of integrating them into optical logic circuits and interfacing them with traditional silicon electronics. One of the most promising types of solid-state nanocrystal QEs are semiconductor quantum dots (QDs) and perovskite nanocrystals (PNCs). Due to the high photoluminescence (PL) quantum yield (QY) they can serve as on-demand sources of single photons, and they can also emit photon pair.[10–12] Emitting a photon pair by a semiconductor QE is a result of the cascade recombination of the excited biexciton state to the ground state via the single-exciton state and the emitted photon pair can be quantum entangled.[4,9,13–16] However, the low PL QY of the biexciton emission in semiconductor QEs strongly limits their use as the source of entangled photon pairs. Due to the strong Coulomb interaction of hot carriers (two holes and two electrons) forming the biexciton, the Auger-like recombination process in nanocrystals is very fast.[17] Thus, nonradiative relaxation from the biexciton state to the single exciton state is the most probable mechanism of the biexciton-to-exciton transition, and the PL QY of this transition is very low compared to the PL QY of the transition of the single exciton to the ground state. The light-matter coupling of QEs with optical microcavities or plasmonic nanocavities can help overcoming this constraint.[9,18,19]

The Purcell effect leads to an increase in the QE operating rate,[20] indistinguishability of the emitted single photons,[21] and entanglement fidelity of the emitted photon pairs.[15] Two types of cavities can be used for obtaining light-matter coupling: optical microcavities (Fabry–Pérot cavities, photonic crystals, Mie resonators, etc.) and plasmonic metal "nanocavities" (single nanoparticles, dimer antennas, nanopatch antennas, metasurfaces, etc.). Optical microcavities offer high values of quality factor, but mode volumes are within the diffraction limit, which limits the absolute values of coupling strength and Purcell factor.[22,23] In contrast, plasmonic nanocavities are metal nanostructures, in which resonant oscillations of the electron density (called localized surface plasmons) occur due to the interaction with light.[24] These oscillations allow much better localization of electromagnetic modes on the nanometer scale, and the maximum possible Purcell factors of plasmon nanostructures can be two orders of magnitude higher than those of optical microcavities.[25] Therefore, a significant acceleration of radiative relaxation makes it possible to achieve a highly efficient on-demand emission of photon pairs from a single semiconductor QE.[26–28] The additional advantage of plasmonic structures is the possibility of using of plasmonic metasurfaces instead of single plasmonic nanoparticles which allows to establish the interaction with macroscopic ensembles of molecules.[29,30]

However, in the photonic structures with QEs attached to the plasmonic nanostuctures, QEs with increased biexciton PL QY are multiphoton emission sources and cannot serve as single-photon sources. Here we have demonstrated the possibility of plasmon-induced reversible switching of semiconductor QEs from single- to multiphoton emission by controllable manipulation on the distance between QE and plasmonic metasurface.

**Results**

For the design of the plasmon-controlled solid-state QEs, we used cesium lead-halide $CsPbBr_{2.5}I_{0.5}$ PNCs synthesized using the modified supersaturated recrystallization technique.[31] The size of the PNCs was about 15 nm (Figure 1a), the PL maximum in the hexane solution under the normal conditions was at a wavelength of 530 nm with a full width at half maximum (FWHM) of 20 nm (Figure 1b) and a PL QY of 25%. The corresponding PL decay kinetics can be well-fitted by the three-exponential function (Figure 1c) with an amplitude weighted average lifetime of 11.5 ns. The QEs represented PNCs incorporated into a 15 nm thick PMMA film on the surface of a glass slide. The optical properties of the PNCs changed when they were embedded into the film: the PL peak was shifted to 520 nm and narrowed to an FWHM of 15 nm (Figure 1d), and the PL lifetime was shortened to the average of 6.3 ns (Figure 1c). We can attribute these changes to partial loss of iodine ions by the PNCs during film fabrication. Indeed, iodine ions can migrate within the crystal structure of PNCs.[32] Due to the stress during the deposition procedure, some iodine ions migrating to the PNC surface may have been oxidized into $I_2$ (iodine), causing irreversible changes in the PNC stoichiometry, thus leading to a blue spectral shift and narrowing of the spectrum.[31,32]

At the beginning of each experiment, we initially identified individual PNCs (Figure 1e) by measuring the second-order cross-correlation function $g^{(2)}$. The single QE behavior of the selected PNCs could be proved if the central peak of the $g^{(2)}$ function of a single QE was below 50% of the side peaks (Figure 1f).[33] In addition, we measured the PL intensity-time traces of these individual PNCs, which exhibited the PL blinking behavior typical of semiconductor nanocrystals (Figure 1g).

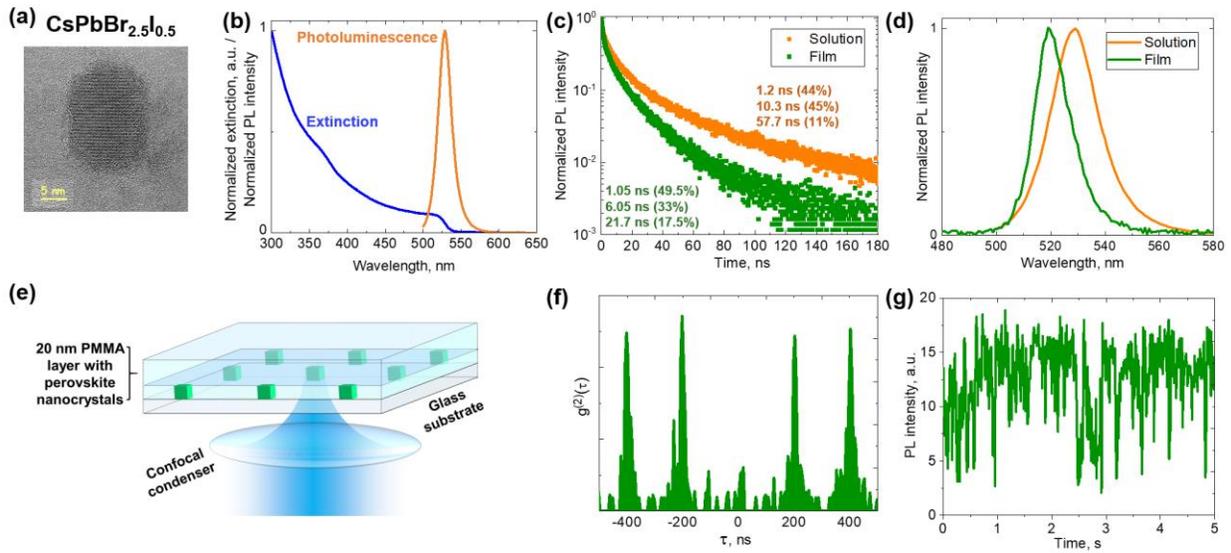

**Figure 1.** Characteristics of the PNCs used. (a) A TEM image of an individual PNC; (b) the extinction and PL spectra of the hexane solution of PNCs; (c) the PL decay kinetics for the hexane solution of PNCs (orange dots) and the PNC–PMMA thin film (green dots); (d) the change in the PL spectrum of PNCs after the transfer from the hexane solution to the thin PMMA film; (e) the measurement geometry; (f) the characteristic $g^{(2)}$ function of an individual PNC; (g) the characteristic PL trace of an individual PNC.

Plasmonic metasurfaces were obtained using PVP-coated 75 nm silver nanocubes (SNCs) commercially available in the form of an ethanol solution from nanoComposix (USA). Their optical properties and SEM images are shown in Figure S3a. Initially, we used the deposition of SNCs directly onto the surface of the PNC–PMMA film by drop-casting the stock solution (Figure S3c) as in previous studies.[26,27] However, the total PNC PL intensity was considerably reduced after treatment of PNCs with ethanol, and the addition of SNCs only slightly increased the PNC PL intensity, not restoring it to the initial value (Figure S4). The drop of the PL intensity due to PNC deterioration in the presence of ethanol did not allow us to discern the PL-enhancing effect of SNCs on the biexciton emission efficiency. Therefore, we changed the approach to the formation of the PNC–SNC hybrid system. To achieve a greater PL enhancement, we replaced the SNC deposition by drop casting with mechanical merging of the thin PNC film and a curved glass surface (with a curvature radius of 50 cm) coated with SNCs (Figure 3a). The extinction spectrum of the SNC plasmonic metasurface had almost the same maximum as in the stock solution (at 515 nm, Figure 3b), which was further supported by numerical simulation of the extinction spectrum of the SNC array on the glass surface (red squares in Figure S3c).

Before the formation of the hybrid system, we identified individual PNCs in the thin PNC–PMMA film using photon correlation spectroscopy (Figure 2c). Then, we positioned the SNC-covered curved plasmonic metasurface on the PNC–PMMA sample. The system's geometry ensured that the distance between the plasmonic metasurface and the surface of the PNC–PMMA film was maintained at less than 30 nm within the area of approximately 0.2 mm$^2$ (see section 2.3 and Figure S5 in the Supplementary Material for details). As shown in previous studies, this interaction distance proved to be adequate in facilitating efficient plasmon-exciton interactions.[26,27] It should be noted that the lateral position of the PNCs in the image after the positioning of the plasmonic metasurface was not changed significantly, and we were able to find the same individual PNCs before and after the interaction with the metasurface (Figures S7 and S9). During the interaction of single PNCs with the metasurface, the PNC PL decay time was considerably shortened (Figure 2i). The $g^{(2)}$ functions of the PNCs were altered, not corresponding to the single-photon emission behavior anymore (Figure 2c). Thus, an increase of the central peak to a value higher than 50% of that of the side peaks evidenced the switching of the QEs to the multiphoton emission mode (Figure 2d). After moving the plasmonic metasurface apart from the PNCs, we were able to study the same single PNCs, and we found that the $g^{(2)}$ function was almost restored to the shape characteristic of the single-photon emission mode and the PL lifetime became close to value measured before the establishment of the plasmon-exciton interaction (Figures 2e and 2i, respectively).

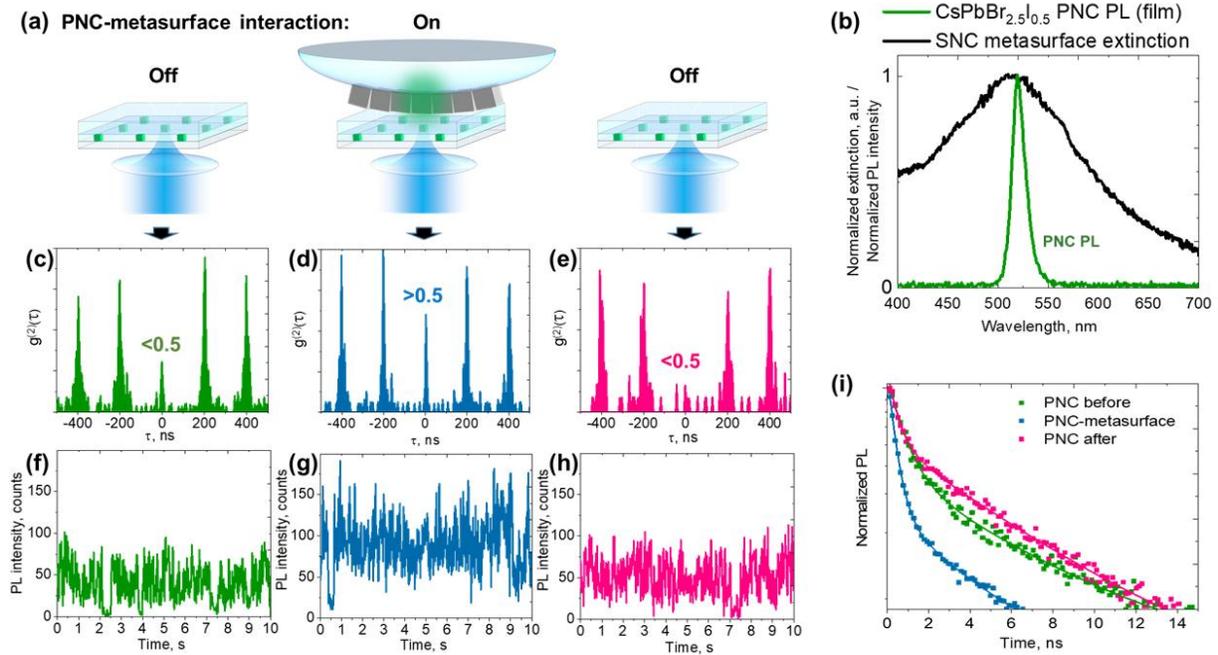

**Figure 2.** Reversible effect of the interaction with the SNC plasmonic metasurface on the emission characteristics of individual PNCs. (a) Schematic representation of the experiment. (b) The extinction spectrum of the SNC plasmonic metasurface (black line), the PNC emission spectrum (green line)(c–e) The g$^{(2)}$ functions of the same single PNC before (c), during (d), and after (e) the interaction with the SNC plasmonic metasurface. (f–h) The PL intensity traces of the same single PNC before (f), during (g), and after (h) the interaction with the SNC plasmonic metasurface. (i) The PL decay curves of the same single PNC before (green), during (blue), and after (pink) the interaction with the SNC plasmonic metasurface.

It is noteworthy that the utilization of the plasmonic metasurface, rather than the deposition of SNCs from a solution or employing techniques for obtaining the hybrid PNC–SNC system had no effect on either morphology and internal structure or chemical composition of PNCs. Therefore, we could restore the initial properties PNCs after elimination of the plasmon–exciton interaction. The convergence of the SNC plasmonic metasurface and PNCs induced an increase in the PL intensity of individual PNCs (Figures 2f–2h, S6). However, after the elimination of the plasmon–exciton interaction, the PL was restored to the initial value, which indicated that the observed PL enhancement can be attributed to the coupling with the plasmonic metasurface.

The observed enhancement of the QE emission may have resulted not only from the Purcell effect, but also from the excitation enhancement caused by PNCs.[27,34] The essential condition for the excitation enhancement is a spectral overlap between the excitation band (peaking at 485 nm in our primary experiments) and the plasmon resonance. This condition was met in our experiments with the metasurface-based hybrid system (Figure 2b).[34] To understand the possible influence of the excitation enhancement on the obtained results, we performed a control experiment using an excitation wavelength of 405 nm. In this case, the excitation was detuned from the plasmon resonance of the SNC plasmonic metasurface. However, the same effects were observed for individual PNCs interacting with the SNC plasmonic metasurface as in the case of the excitation at 485 nm (Figure S7). That is, the spectral overlap between the excitation band and plasmon resonance had no effect on the results. In our experiments, the total plasmon-induced emission enhancement factor for individual PNCs was between 2.2 and 7.5. In addition, we achieved a decrease in the PL lifetime of 1.6 to 2.1 in our hybrid system, although the expected factor of the radiative rate amplification calculated using Equation S1 was in the range from 5 to 12. It should be noted that, for the initially low-emitting PNCs, the PL intensity enhancement was higher than for initially high-emitting PNCs (7.5 times versus 2.2 times, as shown in Figures S7 and 2, respectively). This can be explained if we assume that the radiative rate increase affected the final PL QY stronger in PNCs with a lower initial PL QY. The other reason for the difference in the PL intensity enhancement is the different positions of the QEs relative to the individual SNCs in the metasurface. The calculated Purcell factor for the interaction of the SNCs with the QEs emitting at 520 nm (the spectral maximum of the PNC emission) located at a distance of 7.5 nm from the SNC corner was 8.3; however, the Purcell factor for the PNCs located close to the center of an SNC facet was only 1.3. Moreover, according to the simulated spatial distribution of the SNC plasmon modes (Figure S3d), the area of the possible interaction with an SNC corner is much smaller than the area of the interaction with a flat facet surface. Because the PNCs were located randomly relative to the SNCs, their location close to the flat surface of the SNC facet rather than the SNC corner was more probable. For the ensemble of numerous PNCs interacting with the SNC plasmonic metasurface, the total emission enhancement factor was about 2 (Figure 3), which reflected the effects on both initially low-emitting and initially high-emitting PNCs at all possible positions relative to the SNC corners.

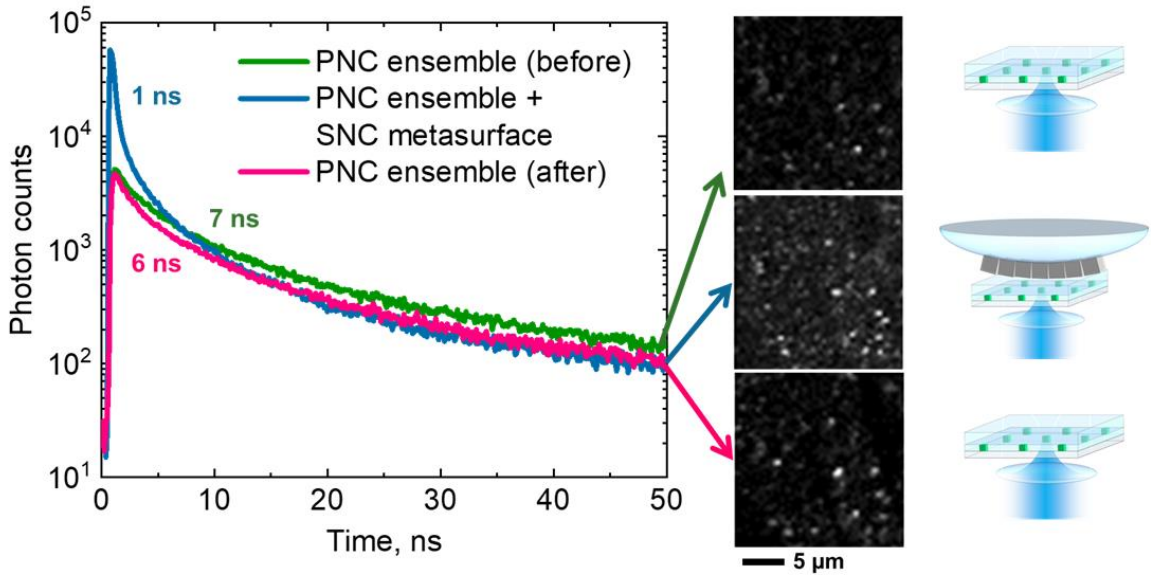

**Figure 3.** PL signals collected from the ensemble of PNCs before the interaction with SNCs (green), during the interaction (blue), and after the removal of the SNC plasmonic metasurface (pink).

Furthermore, we have performed the second control experiment in which we changed the spectral overlap between PNC emission and plasmon resonant peak to investigate whether the reversible interaction will take place in out of the resonance conditions. For this purpose we covered the curved glass surface with plasmon nanoparticles (PNPs) having plasmon resonance at 455 nm wavelength (for details see Section 3 and Figure S8 in the Supplementary Material). Unlike SNCs, which have resonance at the same wavelength as PNC emission, PNP based plasmonic metasurface have resonance exclusively with the excitation wavelength. As a result, we observed only a drop in the PL intensity signal and no changes in other properties under study. This also proves that excitation enhancement did not play a significant role in the designed technique.

To further explore the role of the Purcell effect in the designed technique, we investigated how the overlap between the plasmon resonance and QE emission affected the changes in the PL characteristics of other types of semiconductor QEs. Lead halide perovskites emitting in the red region of the optical spectrum are unstable because of the presence of iodine ions in the crystal structure.[35] Indeed, when we attempted to incorporate red-emitting PNCs into the PMMA thin film using spin-coating, this resulted in destruction of the red-emitting PNCs. To mitigate this problem, we used CdSe-based core/shell QDs (QD560 and QD620 samples; see Section 2.2 of the Supplementary Material for details) with emission wavelengths that are not in resonance with the SNC plasmonic metasurface, which weakened the Purcell effect (Figures 4a, 4b).

In the case of the hybrid system based on the QD560 sample, the overlap between the QD emission spectrum and the plasmon resonance of the SNC plasmonic metasurface was smaller than in the case of PNCs (Figure 4b). In this experiment, the PL lifetime was considerably shortened (Figure 4e) and the PL intensity was increased by 30% (Figure 4c). This increase was smaller than that observed in the case of PNCs because of the lower Purcell factor. Indeed, as can be seen from Figure 4a, the calculated Purcell factor for the case when the QE is near the center of the SNC facet is only 0.7 at the wavelength corresponding to the QD560 emission maximum (Figure 4a), which can be the reason of the relatively low average enhancement of the PL intensity. As in the case of PNCs, we observed switching of the emission of individual QDs in the QD560–metasurface hybrid to the multiphoton mode as result of interaction with the SNCs and then back to the single photon mode after the removal of the metasurface (Figure 4d).

In contrast, plasmon–exciton interaction in the case of red-emitting QDs from the QD620 sample led to a strong quenching of the PL of individual QDs accompanied by shortening of the PL lifetime (Figures 4f, 4g). The calculated Purcell factor for the wavelength of 620 nm (the QD620 emission maximum) at the center of the SNC facet was 0.2, which means that fivefold decrease in the PL intensity can be expected. The observed quenching factor was smaller than the estimated value (by a factor of about four) because the Purcell factor was still high at the SNC corners (6.2, see Figure 4a); however, the contribution of the SNC corner area to the total area of the QE–metasurface interaction was rather low. Moreover, we did not observe any effect of the metasurface on the $g^{(2)}$ function (Figure 4h) of individual QDs, which supports the assumption that the increase in the radiative recombination rate (Purcell effect) led to the enhancement of biexciton emission and levelling of the exciton and biexciton PL QYs in the hybrid systems characterized by a large overlap between the emission maximum and plasmon resonance. Thus, we assume that the Purcell effect was the main factor that allowed us to reversibly switch the QEs between the single-photon and the enhanced photon pair emission modes.

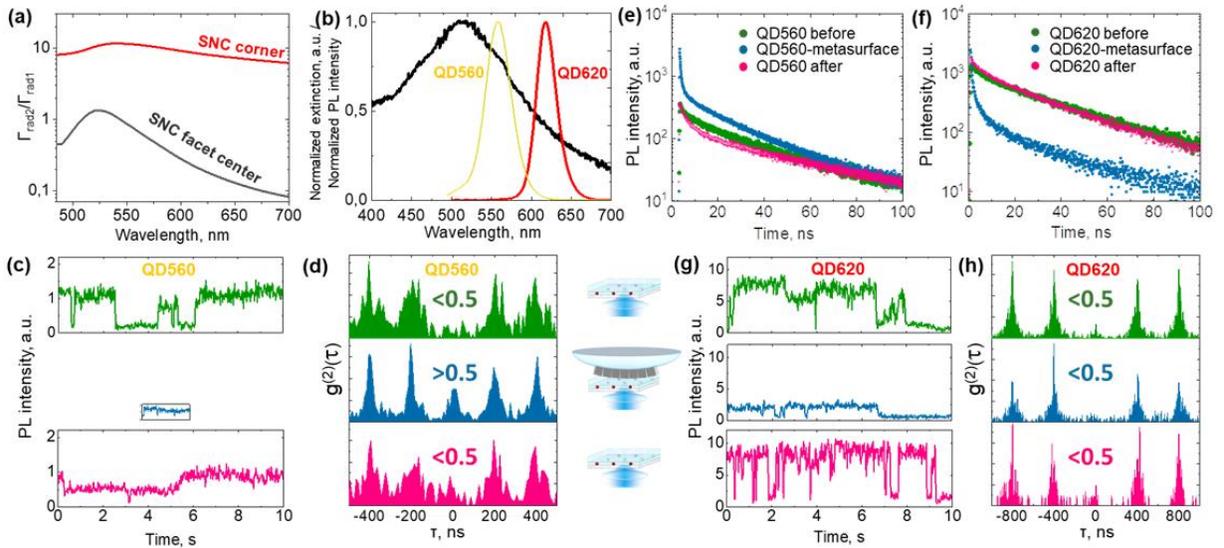

**Figure 4.** Effect of the SNC plasmonic metasurface on the PL of individual QDs. (a) The calculated acceleration of the QE radiative rate 7.5 nm away from the center of the SNC facet (black line) and from the SNC corner (red line). (b) The PL spectra of the QD560 (yellow line) and QD620 (red line) samples compared to the extinction of the SNC plasmonic metasurface

(black line). (c) The PL intensity trace and (d) second-order cross-correlation function $g^{(2)}$ of the same single QD (QD560 sample) before the interaction with the SNC plasmonic metasurface (green), during the interaction (blue), and after the removal of the metasurface (pink). (e, f) The PL decay kinetics of the same single QDs from the QD560 (e) and QD620 (f) samples before the interaction with the SNC plasmonic metasurface (green dots), during the interaction (blue dots), and after the removal of the metasurface (pink dots). (g) The PL intensity trace and (h) second-order cross-correlation function $g^{(2)}$ of the same single QD (QD620 sample) before the interaction with the SNC plasmonic metasurface (green), during the interaction (blue), and after the removal of the metasurface (pink).

**Summary and outlook**

We have designed a plasmon–exciton hybrid system with a controlled Purcell factor based on disordered plasmonic metasurfaces and semiconductor QEs (semiconductor QDs and PNCs). This system allowed us to reversibly change (increase or decrease) the intensity of the emission and shorten the emission lifetime for an ensemble of QEs of different types. Furthermore, for individual QEs, we were able to switch reversibly the emission mode from the single-photon to the photon pair. In contradistinction to previous works, the proposed technique does not require the precise positioning of the AFM tip to the QE.[36,37] Instead, we affect the total array of QEs in the same way, and for choosing the necessary QE only the positioning of the objective is required. The further development of the proposed technique could involve the creating an array of nano-patch antennas, similar to demonstrated in the work of Dhawan et al. but with a controlled gap distance.[28] Moreover, precise positioning of QEs in hotspots of plasmonic surface will enable an increase in the Purcell factor, leading to more efficient modulation of the emission mode. This concept can be used for the production of arrays of QEs with controlled emission characteristics, including the single-photon/multiphoton mode, emission rate, and intensity. Further optimization of the structure of the plasmon and exciton arrays will make it possible to reach higher values of the Purcell factor and, hence, more efficiently modulate the emission mode. The designed system can be used for designing both light-controlled QEs for quantum information applications and QLEDs with switchable emission characteristics controlled by the light–matter coupling phenomenon.

## A. Materials and methods

*1. Experimental setup*

For the measuring of the PL signal and photon correlation spectroscopy we used MicroTime 200 inverted fluorescent microscope from PicoQuant, Germany (Figure S1). We used water immersion ×60 objective with 1.2 NA. For the excitation we used picosecond lasers with 485 nm or 405 nm wavelengths and pulse duration of about 200 ps. For the photon correlation spectroscopy experiments we used Hunburry-Brown-Twiss geometry of the setup with two avalanche photodetectors. To cut the excitation line wavelength we used long-pass optical filter with 500 nm cutting wavelength. We also used 100 μm pinhole to purify the PL signal.

*2. Synthesis of $CsPbBr_{2.5}I_{0.5}$ perovskite nanocrystals*

For the synthesis of $CsPbBr_{2.5}I_{0.5}$ perovskite nanocrystals (PNCs) all the reagents were purchased from Sigma Aldrich and were used directly without further purification. First, 20 mg of CsBr, 26 mg of $PbBr_2$, and 11 mg of $PbI_2$ were dissolved in 2.235 mL of DMF. 223 μL of OA and 112 μL of OAm were added to stabilize the precursor solution. Then, 20 μL of the precursor solution was quickly added into 2 mL of toluene under vigorous stirring at 2000 rpm. Immediately after the formation of the green colored solution it was transfused to 10 mL of n-hexane. The optical density of the final PNC at first exciton peak was 0.1. All above operations were implemented at normal conditions.

*3. Synthesis of quantum dots*

CdSe/ZnS/CdS/ZnS core-multishell quantum dots (QD560 sample) were synthesized according to the procedure described in our previous study.[27] The method for synthesizing CdSe/CdS QDs (QD620 sample) is based on the results of the work of Bawendi and coworkers [38]. See Section 2.1 in the Supplementary Material for more details.

*4. Preparation of quantum emitters in poly(methyl methacrylate) film samples*

Previously we reported that the best PL enhancement was achieved when the distance between QDs and plasmonic nanoparticles was in-between 10 and 20 nm [27,39]. Thus, in this study we prepared QD/PNC arrays in poly(methyl methacrylate) (PMMA) films of the total thickness of approximately 15 nm. Thin films were deposited by spin-coating method using a Model KW-4A Spin Coater at 1000 RPM during 60 s cycles for each layer. Firstly, the PMMA layer was deposited on a previously cleaned glass substrate using 70 uL of 0.5 wt% PMMA (120 000 molecular weight) solution in toluene was deposited onto the glass surface. Then the QDs (or PNCs) were deposited onto the PMMA surface from a hexane solution with a molar concentration of about $10^{-6}$ M. The second PMMA layer was deposited in the same way as the first one.

*5. Preparation of the perovskite nanocrystal samples covered with silver plasmonic metasurface*

For the first experiments in which the interaction of PNCs with SNC covered with 3 nm of Polyvinylpyrrolidone (PVP) coating (SEM image presented at Figure S3a) was studied, we used the deposition of 2 μL of the SNC ethanol solution (purachsed from nanoComposix) on the top of the PNC-PMMA thin films (Figure S3b).

For the "dry" reversible deposition of the SNC metasurface on the surface of QE-PMMA thin films we used a 2.5 cm round glass curved surface with 50 cm curvature radius. Before the deposition we dropped 2 µl of the stock solution of SNCs in the center of the curved glass surface and then waited till ethanol was evaporated and the array of SNC was formed on the surface of the curved glass.

To change the plasmon resonance spectral position we attached SNCs to the curved glass surface using different deposition method. Control sample, prepared by polyelectrolyte-assisted layer-by-layer assembly method.[40,41] exhibit an absorption spectrum changed in comparison to the drop-casting deposition. First, the curved glass surface was washed with alkaline cleaning solution and sonicated in water and then in ethanol. Clean substrate was then immersed in freshly prepared Poly(diallyldimethylammonium chloride) solution purchased from Sigma-Aldrich (1mg/mL in 0.5M NaCl) for 20 min and rinsed with MiliQ water. Subsequently, the curved glass was immersed for 1 hour in ethanolic solution of PVP coated silver nanocubes ($[Ag^0] = 0.6$mM). Substrate was rinsed with MiliQ water and dried under argon flow. Due to the chemical degradation and etching of the SNC corners during the deposition the formed layer of PNPs was corresponding to the layer of near-spherical plasmon nanoparticles of about 75 nm size with the plasmon resonance at 455 nm wavelength (Figure S8a).[42]

**Supplementary Information**

Details about the experimental setup, nanocrystal synthesis and characterization, thin-film sample preparation, deposition of SNCs onto the surface of the PNC–PMMA film, and related experimental results; the description of the preparation and use of the nanocrystal-coated curved glass surface; results of photoluminescence and photon correlation spectroscopies at an excitation wavelength of 405 nm using PNPs and SNCs and at an excitation wavelength of 485 nm using SNCs; details of the Purcell factor calculation.

**Acknowledgments and funding**

The study was funded by the European Union's Horizon 2020 research and innovation programme under the Marie Sklodowska-Curie, grant agreement no. 101025664 (QESPEM). YR and VK acknowledge support by MCIN and by the European Union NextGenerationEU/PRTR-C17.I1, as well as by IKUR Strategy under the collaboration agreement between Ikerbasque Foundation and Material Physics Center on behalf of the Department of Education of the Basque Government. The part of the study related to the synthesis of semiconductor quantum dots was supported by the Russian Science Foundation (RSF), grant no. 21-79-30048. YR and AO also acknowledge support from the Office of Naval Research Global (Award No. N62909-22-1-2031). IN acknowledges support from the French National Research Agency ANR, grant no. ANR-20-CE19-009-02, and from the Université de Reims Champagne-Ardenne.

**Conflict of interest**

The authors have no conflicts to disclose.

**Data availability**

All source data that support the plots within this paper and other findings of this study are available from the corresponding authors upon reasonable request.

# SUPPLEMENTARY MATERIAL

## Table of contents

**1. Experimental setup**

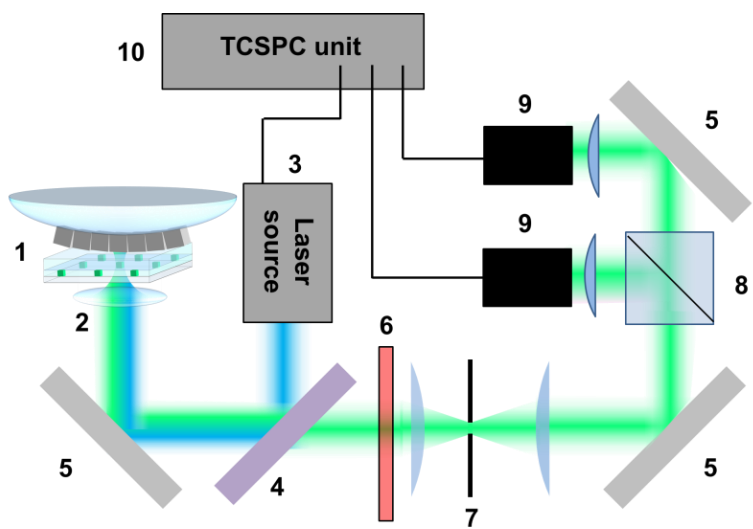

**Figure S1.** Experimental setup (fluorescent microscope MicroTime 200 from PicoQuant) for time-correlated single photon counting measurements, including cross-correlation measurements. 1 – sample; 2 – water-immersion objective (×60, NA=1.2); 3 – picosecond laser source (485 nm or 405 nm wavelength); 4 – dichroic mirror; 5 – mirrors; 6 – 500 nm wavelength long-pass

optical filter; 7 – 100 µm pinhole; 8 – 50/50 beam-splitter; 9 – avalanche photodetectors; 10 – TCSPC unit.

## 2. Synthesis and preparation of samples

*2.1 Synthesis of quantum dots*

CdSe/ZnS/CdS/ZnS core-multishell quantum dots (QD560 sample) were synthesized according to the procedure described in our previous study.[S1] CdSe nanocrystals with diameter 2.3 were obtained by the hot injection procedure using Cd hexadecylphosphonate and TOP-Se precursors at 240°C, purified using several sedimentation/redispersion cycles and gel permeation chromatography, and coated with a multilayer ZnS/CdS/ZnS shell in a layer-by-layer SILAR routine.[S2] The finally obtained quantum dots (QDs) were treated with hexadecylammonium palmitate and dried in TOPO matrix to obtain a solidified sample for further use. Their extinction and PL spectra as well as PL decay kinetics are presented in Figure S2 by yellow lines. PL quantum yield (QY) in the hexane solution was about 80%, and about 50-60% in the PMMA film.

The method for synthesizing CdSe/CdS QDs (QD620 sample) is based on the results of the work of Bawendi and coworkers.[S3] The synthesis of CdSe cores with a diameter of about 3.5 nm was carried out by colloidal synthesis in an organic medium as described in detail in the previous study.[S4] Next, a thick shell of CdS was deposited on the CdSe nuclei prepared by the method described above, the surface of which was coated by OLA ligands, as follows. A solution of purified CdSe nuclei in toluene was transferred into a mixture of 12 ml of ODE and 12 ml of OLA in a three-necked flask, after which the toluene evaporated under vacuum in temperature of 60-100°C. Next, removal of dissolved gasses from the mixture was assured by stirring the solution under vacuum for 20 min, after which heating of the reaction mixture to 310°C at a rate of 20 K/min in an argon atmosphere was started. When the temperature reached 240°C, the introduction of prefabricated solutions of cadmium oleate (0.4 M, with a 3-fold excess of oleic acid relative to the cadmium introduced into the reaction) in ODE and n-octanethiol (0.4 M) was injected into the flask using programmable syringe pumps at a rate corresponding to the addition of the reagents sufficient to form 2 shell monolayers in 1 h. After the last portions of the precursors were added, additional 2.3 ml of oleic acid was introduced into the flask. From that moment the mixture was kept at 310°C for 1 h. After reaction mixture was gradually cooled down, nanocrystals were precipitated using methyl acetate and redispersed in 10 mL of hexane. Subsequently, 400 mg of of tri-n-octylphosphine oxide was added to the solution, it was transferred to a crystallizer, and dried to obtain a dry powder. The TEM image of the QD620 sample is presented in Figure S2a. Its extinction, PL spectra, and PL decay kinetics are presented in Figures S2b-d by red lines. PL quantum yield (QY) in the hexane solution was about 40%, and about 30% in the PMMA film.

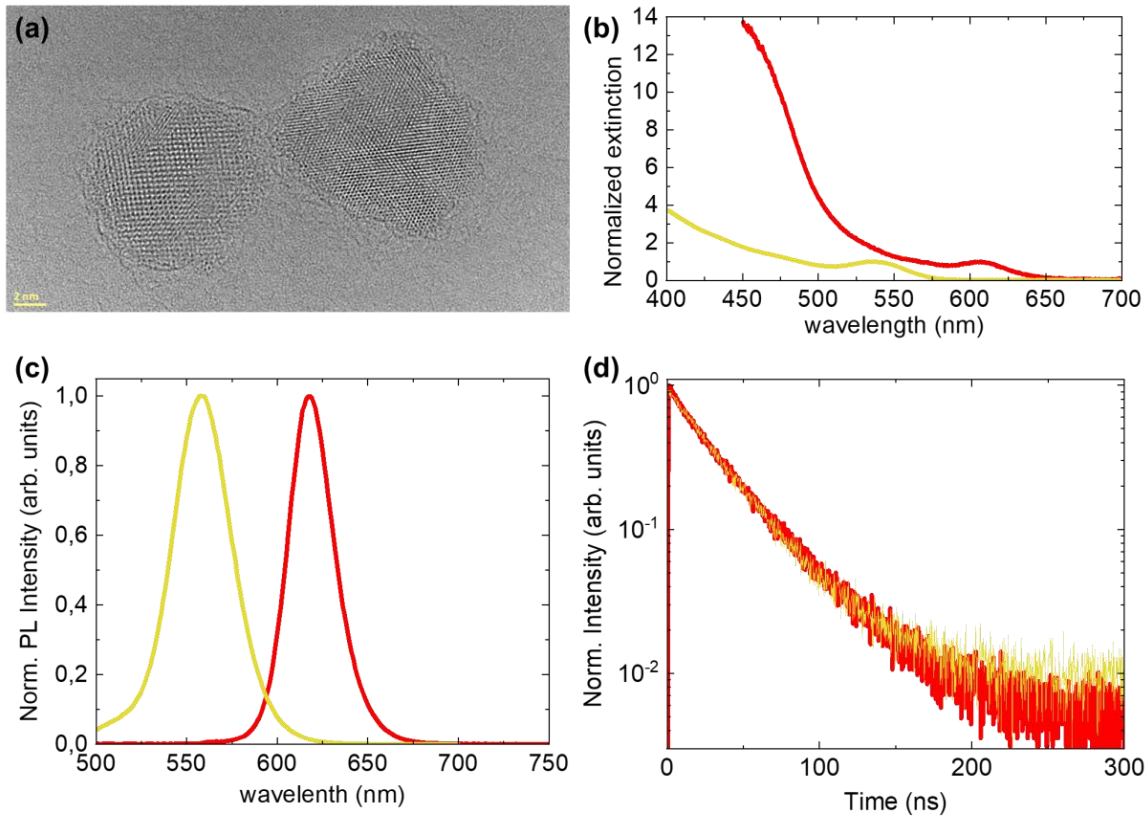

**Figure S2.** Properties of used QD samples. (a) TEM image of QD620. (b) Extinction spectra of hexane solutions of QD560 (yellow) and QD620 (red). (c) PL spectra of hexane solutions of QD560 (yellow) and QD620 (red). (d) PL decay kinetics of hexane solutions of QD560 (yellow) and QD620 (red).

*2.2 Preparation of the perovskite nanocrystal samples covered with silver nanocubes*

For the first experiments in which the interaction of PNCs with SNC covered with 3 nm of Polyvinylpyrrolidone (PVP) coating (SEM image presented at Figure S3a) was studied, we used the deposition of 2 µL of the SNC ethanol solution (purachsed from nanoComposix) on the top of the PNC-PMMA thin films (Figure S3b). The extinction spectrum of the SNC film on the top of PMMA one can see in Figure S3c. Compared with the SNC solution spectrum the only sufficient change was a shift of the extinction maximum from 510 nm to 515 nm (Figure S3c). The FEM modelled extinction spectrum of the array of SNCs coated with 3 nm of PVP on the top of PMMA film (Figure S3d) is also in good agreement with experimental one (Fgiure S3c) which confirms the absence of the strong aggregation of SNCs on the top of PMMA.

In this approach, the total PL intensity was notably reduced following the exposure of PNCs to ethanol. Furthermore, the introduction of SNCs provided only a slight restoration of the initial PL value (figure S4). As the quantity of SNCs increased, the signal quality improved, resulting in a maximum PL intensity of up to 25% of the initial value (figure S4l). At the same time, the $g^{(2)}$ function of initially single PNC changed to the behavior typical for multiphoton emission with central peak of ~90% of the side peaks (figure S4k). Throughout all of our experiments, the excitation energy remained below 120 fJ per pulse which ensured the average exciton occupancy below 0.1. Thus, the increase of the intensity of the central peak of $g^{(2)}$ function can be explained as levelling of the exciton and biexciton PL QYs. Meeting the spectral criteria with the emission peak of PNC at 520 nm, we effectively satisfied the necessary conditions for leveraging the Purcell effect. Consequently, both exciton and biexciton photoluminescence quantum yield (PL QY) experienced significant enhancement.

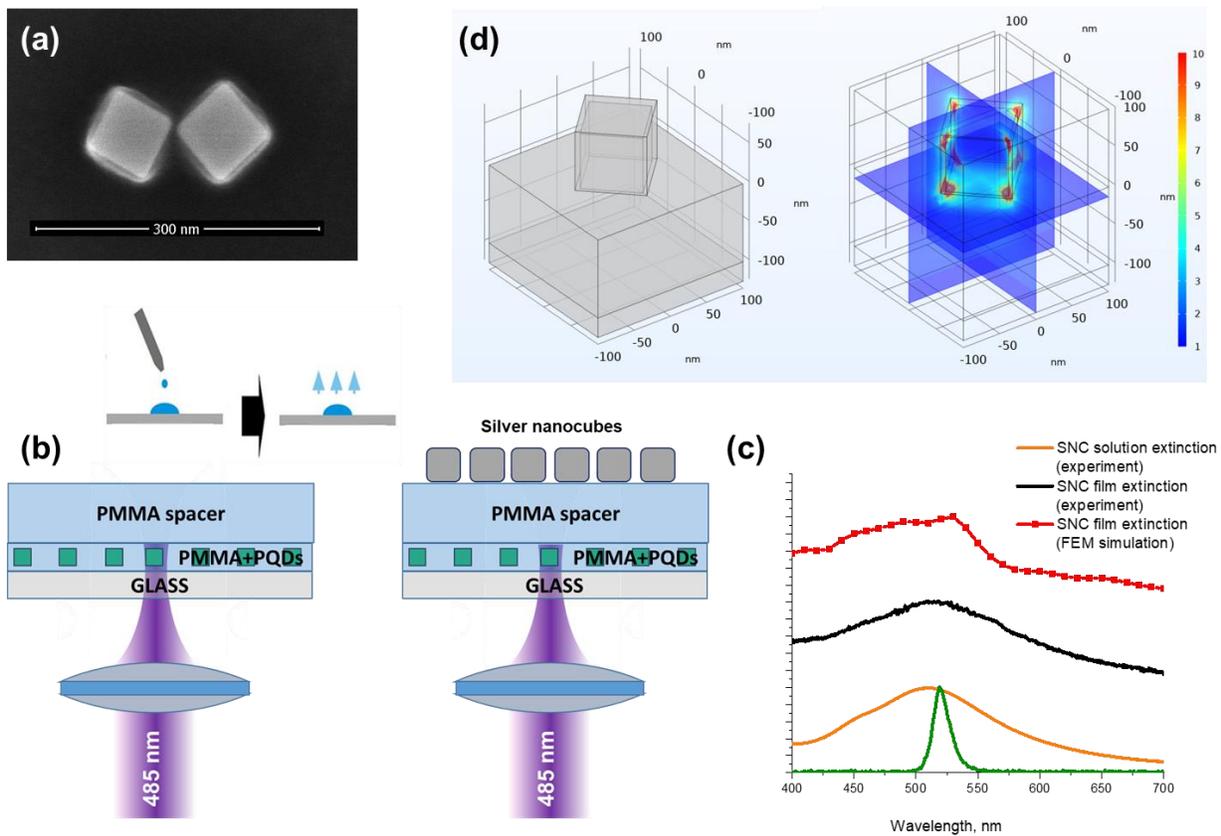

**Figure S3.** (a) SEM image of silver nanocubes. (b) FEM model of the SNC metasurface on the surface of glass/PMMA. (c) Schematic representation of the experiment with a deposition of SNCs on the PNC-PMMA film. (d) Experimental SNC film spectra in solution (orange line) and in the form of a metasurface (black line) as well as calculated spectrum of the SNC array on glass/PMMA film (red dots).

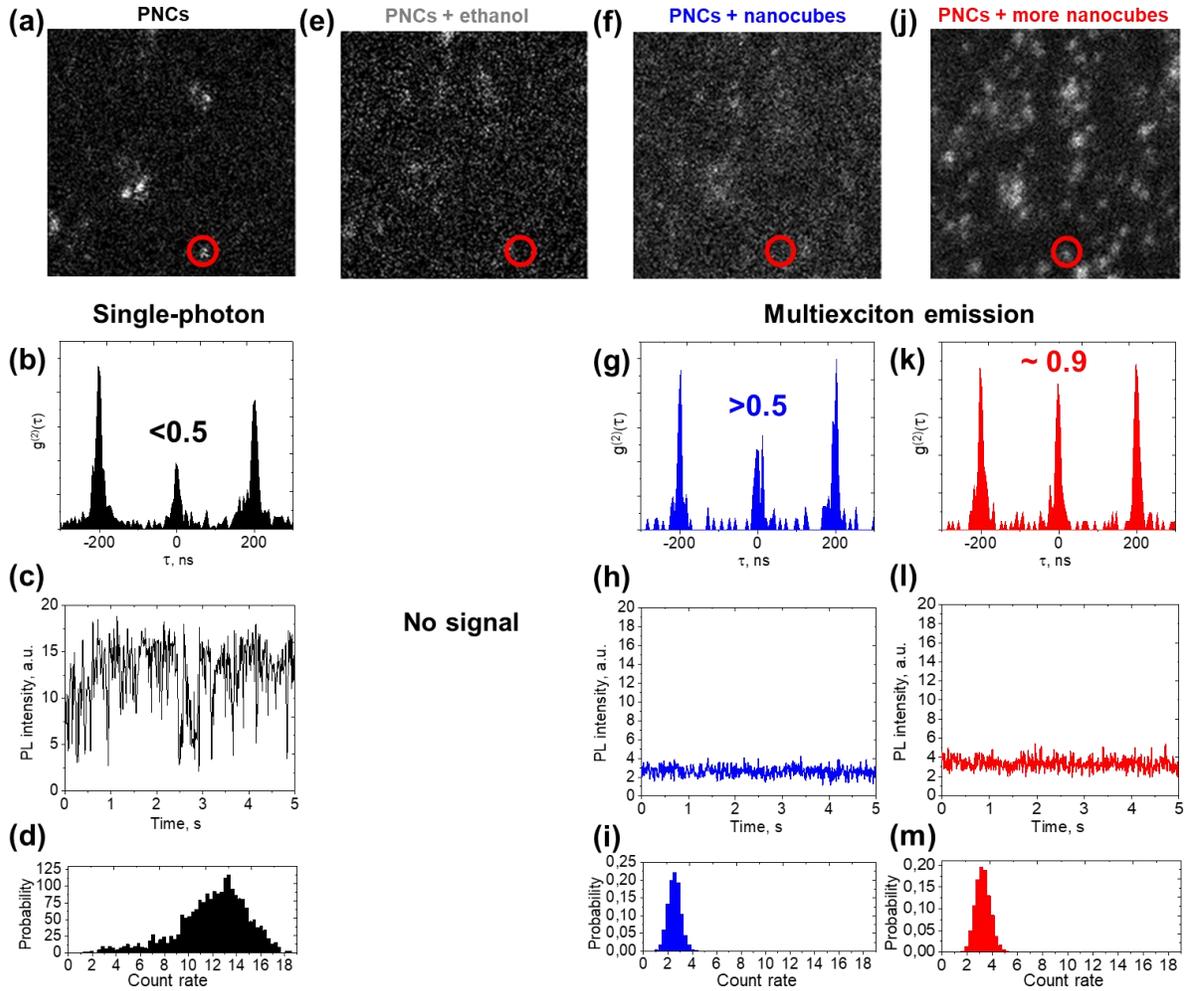

**Figure S4.** Single PNC measurements (a-d) before the deposition of SNC layer, (e) after deposition of ethanol, (f-i) after the deposition of first drop of SNC solution, and (j-m) after the deposition of three drops of SNC solution. (a), (e), (f), (j) – PL images; (b), (g), (k) – second order cross-correlation $g^{(2)}$ function; (c), (h), (l) – PL traces; (d), (i), (m) – PL intensity distribution.

*2.3 Preparation of the SNP and PNP curved plasmonic metasurfaces and their deposition on the QE-PMMA samples*

For the "dry" reversible deposition of the silver nanocubes (SNCs) array on the surface of QE-PMMA thin films we used a 2.5 cm round glass curved surface with 50 cm curvature radius. Before the deposition we dropped 2 μl of the stock solution of SNCs in the center of the curved glass surface and then waited till ethanol was evaporated and the layer of SNC was formed on the surface of the curved glass. Then we repeated the procedure two more times. The spectrum of SNC film one can see in the figure 2b of the main manuscript. We assume the absence of the aggregation of SNCs since the plasmon spectrum was only slightly shifted from 510 nm to 515 nm wavelength and slightly broadened.

As a control experiment we have measured a sample of SNCs attached to the curved glass surface using different deposition method. Control sample, prepared by polyelectrolyte-assisted layer-by-layer assembly method,[S5,S6] exhibit an absorption spectrum changed in comparison to the drop-casting deposition. First, the curved glass surface was washed with alkaline cleaning solution and sonicated in water and then in ethanol. Clean substrate was then immersed in freshly prepared Poly(diallyldimethylammonium chloride solution purchased from Sigma-Aldrich (1mg/mL in 0.5M NaCl) for 20 min and rinsed with MiliQ water. Subsequently, the curved glass was immersed for 1 hour in ethanolic solution of PVP coated silver nanocubes ([$Ag^0$] = 0.6mM). Substrate was rinsed with MiliQ water and dried under argon flow. Due to the chemical degradation and etching of the SNC corners during the deposition the formed layer of plasmon nanoparticles was corresponding to the layer of near-spherical plasmon nanoparticles of about 75 nm size with the plasmon resonance at 455 nm wavelength (Figure S8a).[S7]

During the experiments, we placed the curved glass surface covered with plasmon nanostructures on the surface of the 0.14 mm glass covered with QE-PMMA film so that the center of the curved glass surface was aligned with the objective focus. Under the weight of the curved glass surface, the glass substrate was bent and we could measure it by changing the z-position of the objective and finding the surface of the glass by interferometry technique. The change of the z-position in the center of the glass substrate was about 50 μm. Considering the curvature of the curved glass surface and the glass substrate (the diameter of the glass substrate able to be bent was 2.5 cm) we calculated the diameter of the area where the distance between the surface of the glass substrate and the surface of the curved glass surface is less than 30 nm (figure S5). We considered this area as the area of the possible interaction between QE and plasmonic layers. The size of this area was about 0.5 mm, which is big enough to be sure that with plasmonic nanostructure layer we cover the same QE we measured before the deposition.

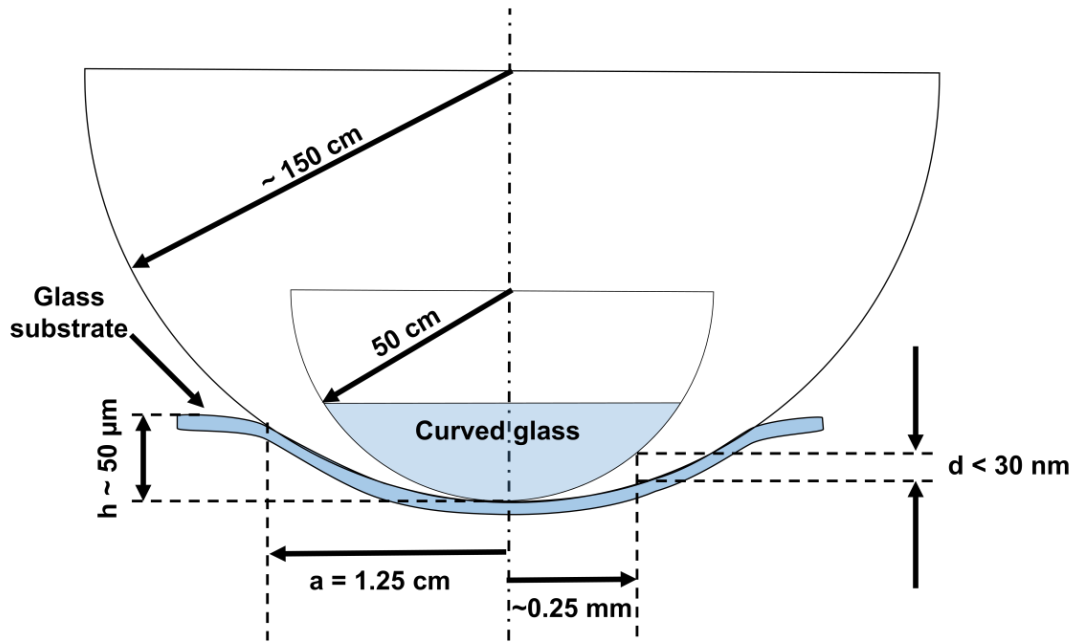

**Figure S5.** Geometry of the deposition of the SNC plasmonic metasurface on the glass substrate covered with the PNC-PMMA film.

## 3. Photoluminescence and photon correlation spectroscopy results at 485 nm and 405 nm excitation wavelengths

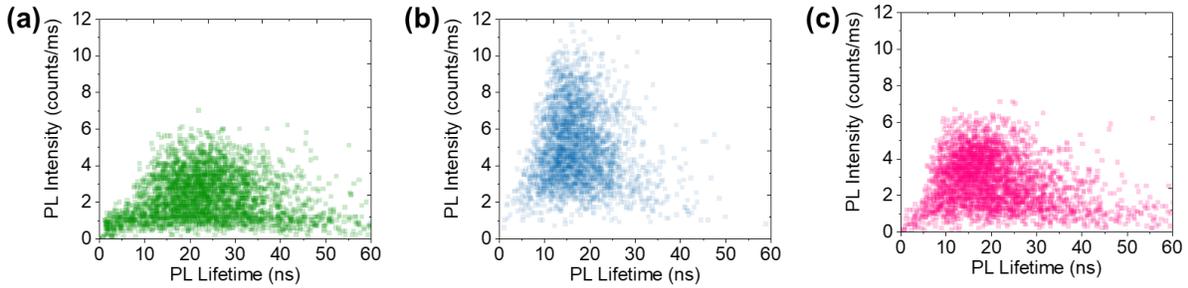

**Figure S6.** Intensity-lifetime distribution of the same single PNC before (a), during (b) and after (c) the interaction with the SNC plasmonic metasurface.

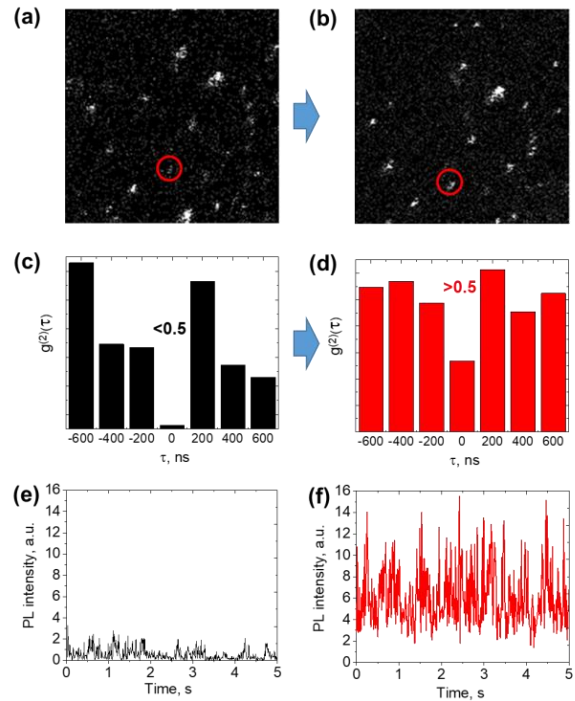

**Figure S7.** PL properties of initially low-emitting PNC in PMMA sample before (a, c, e) and after (b, d, f) the deposition of the SNC plasmonic metasurface under the 405 nm wavelength laser excitation. (a, b) – PL image; (c, d) - second order cross-correlation $g^{(2)}$ function; (e, f) – PL traces.

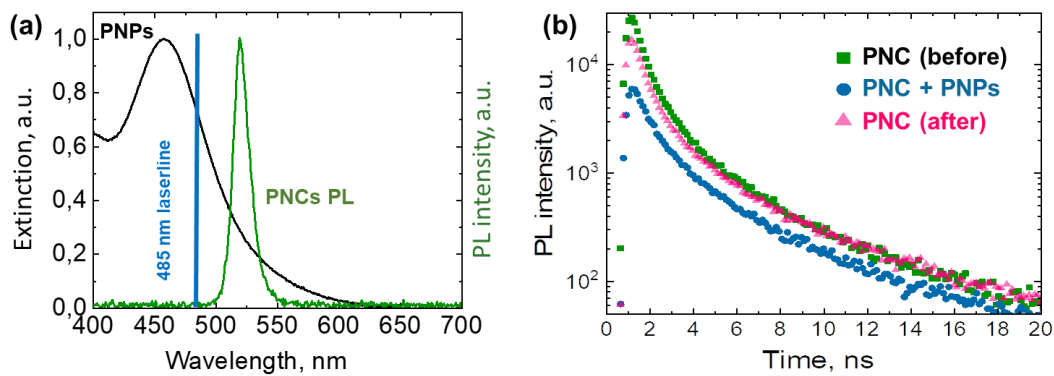

**Figure S8.** (a) Extinction spectrum of the PNP plasmonic metasurface (black line) and PNCs PL spectrum (green line). (b) PL decay kinetics of PNCs before the interaction with the PNP plasmonic metasurface (black squares), during the interaction (red circles), and after the elimination of the interaction (blue triangles).

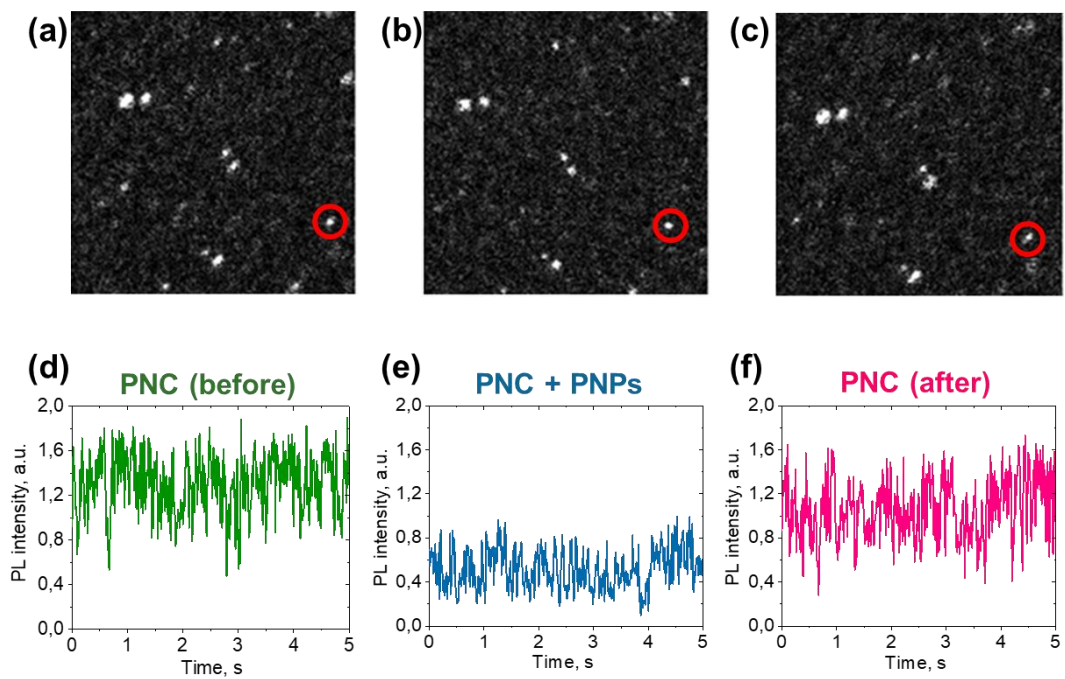

**Figure S9.** Effect of the deposition of the PNP plasmonic metasurface on the emission properties of single PNC. (a-c) PL images of the PNC ensemble (a) before the interaction, (b) during the interaction, and (c) after removing of the PNP plasmonic metasurface. (d-f) PL intensity traces of single PNC (d) before the interaction, (e) during the interaction, and (f) after removing of PNPs.

## 4. Calculation of the effect of the plasmonic metasurface on QE emission

For the calculation of the Purcell factor based on the experimental results we used following equation used in our previous study [S8]:

$$PL_2/PL_1 \approx \frac{F \cdot \Gamma_{rad} + F \cdot \Gamma_{nr}}{F \cdot \Gamma_{rad} + \Gamma_{nr} + \Gamma_{ET}}. \quad (S1)$$

where $PL_1$ and $PL_2$ are PL intensities of QE without effect of plasmons and during the interaction with plasmons correspondingly, $\Gamma_{rad}$ is the radiative rate of the QE, $\Gamma_{nr}$ is the intrinsic nonradiative rate of the QE, $\Gamma_{ET}$ is the rate of the energy transfer from the QE to dark plasmon mode, $F$ is the Purcell factor which we define as the acceleration of the radiative rate $\Gamma_{rad}$ without accounting of the coupling with dark plasmon modes.

## S5. References

S1 V. Krivenkov, D. Dyagileva, P. Samokhvalov, I. Nabiev, and Y. Rakovich, "Effect of Spectral Overlap and Separation Distance on Exciton and Biexciton Quantum Yields and Radiative and Nonradiative Recombination Rates in Quantum Dots Near Plasmon Nanoparticles," Ann. Phys. 532(8), 2000236 (2020).

S2 P. Samokhvalov, P. Linkov, J. Michel, M. Molinari, and I. Nabiev, "Photoluminescence quantum yield of CdSe-ZnS/CdS/ZnS core-multishell quantum dots approaches 100% due to enhancement of charge carrier confinement," Proc. SPIE 8955, 89550S (2014).

S3 O. Chen, J. Zhao, V.P. Chauhan, J. Cui, C. Wong, D.K. Harris, H. Wei, H.-S. Han, D.